\documentclass[sigconf]{acmart}
 \usepackage{amsmath,amssymb,amsfonts,bm}
\usepackage{graphicx}
\usepackage{textcomp}
\usepackage{xcolor}
\usepackage{float}
\usepackage{array}
\usepackage{tabularx}
\usepackage{array}
\usepackage{balance}
\usepackage{marvosym}
\usepackage{multirow}
\usepackage{array}
\usepackage{enumitem}
\usepackage{booktabs}
\usepackage{algpseudocode}
\usepackage[ruled]{algorithm2e} 
\usepackage{xcolor}
\usepackage{pifont}
\usepackage{url}
\usepackage{algpseudocode}
\usepackage{subfig}
\usepackage[ruled]{algorithm2e}
\copyrightyear{2022}
\acmYear{2022}
\setcopyright{acmcopyright}\acmConference[SIGIR '22]{Proceedings of the 45th International ACM SIGIR Conference on Research and Development in Information Retrieval}{July 11--15, 2022}{Madrid, Spain}
\acmBooktitle{Proceedings of the 45th International ACM SIGIR Conference on Research and Development in Information Retrieval (SIGIR '22), July 11--15, 2022, Madrid, Spain}
\acmPrice{15.00}
\acmDOI{10.1145/3477495.3531775}
\acmISBN{978-1-4503-8732-3/22/07}
\begin{document}

\title{On-Device Next-Item Recommendation with \\Self-Supervised Knowledge Distillation}

\author{Xin Xia}
\affiliation{%
	\institution{The University of Queensland}
	\city{Brisbane}
	\country{Australia}}
\email{x.xia@uq.edu.au}

\author{Hongzhi Yin}
\authornote{Corresponding author.}
\affiliation{%
	\institution{The University of Queensland}
	\city{Brisbane}
	\country{Australia}}
\email{h.yin1@uq.edu.au}

\author{Junliang Yu}
\affiliation{%
	\institution{The University of Queensland}	
	\city{Brisbane}
	\country{Australia}}
\email{jl.yu@uq.edu.au}

\author{Qinyong Wang}
\affiliation{%
	\institution{Baidu Inc.}
	\city{Beijing}
	\country{China}}
\email{wangqinyong@baidu.com}

\author{Guandong Xu}
\affiliation{%
	\institution{University of Technology Sydney}
		\city{Sydney}
	\country{Australia}}
\email{Guandong.Xu@uts.edu.au}

\author{Nguyen Quoc Viet Hung}
\affiliation{%
	\institution{Griffith University}
		\city{Gold Coast}
	\country{Australia}}
\email{quocviethung1@gmail.com}
\fancyhead{}

\begin{abstract}
Modern recommender systems operate in a fully server-based fashion. To cater to millions of users, the frequent model maintaining and the high-speed processing for concurrent user requests are required, which comes at the cost of a huge carbon footprint. Meanwhile, users need to upload their behavior data even including the immediate environmental context to the server, raising the public concern about privacy. On-device recommender systems circumvent these two issues with cost-conscious settings and local inference. However, due to the limited memory and computing resources, on-device recommender systems are confronted with two fundamental challenges: (1) how to reduce the size of regular models to fit edge devices? (2) how to retain the original capacity? \par
Previous research mostly adopts tensor decomposition techniques to compress the regular recommendation model with limited compression ratio so as to avoid drastic performance degradation. In this paper, we explore ultra-compact models for next-item recommendation, by loosing the constraint of dimensionality consistency in tensor decomposition. Meanwhile, to compensate for the capacity loss caused by compression, we develop a self-supervised knowledge distillation framework which enables the compressed model (student) to distill the essential information lying in the raw data, and improves the long-tail item recommendation through an \textit{embedding-recombination} strategy with the original model (teacher). The extensive experiments on two benchmarks demonstrate that, with 30x model size reduction, the compressed model almost comes with no accuracy loss, and even outperforms its uncompressed counterpart in most cases.
\end{abstract}



\maketitle

\section{Introduction}
Benefiting from the huge model size and powerful neural architectures, deep learning-based recommender systems show unparalleled capacity in mining users' latent interests \cite{zhang2019deep}. However, these systems are fully constructed at the server side and rely on abundant storage, memory and computing resources in the cloud. Despite their effectiveness, running these systems is at the cost of a huge carbon footprint \cite{himeur2021survey}. The frequent training and update of recommendation models and the high-speed processing for millions of concurrent user request are driven by a large number of energy-consuming CPUs/GPUs, requiring a vast amount of electric power. Meanwhile, deep recommendation models are fueled by massive user behavior data. When serving users, these cloud-based recommender systems even ask for the immediate contextual data for real-time inference, raising public concern about privacy \cite{jeckmans2013privacy}. 
\par
Recently, on-device machine learning \cite{dhar2021survey} has attracted increasing attention, which enables models to run on personal devices. In this learning paradigm, the machine learning model is first trained on the cloud, and then is downloaded and deployed on local energy-saving devices such as smartphones. When using the model, users no longer need to upload the sensitive data to servers, and therefore can enjoy low-latency services that are orders of magnitude faster than server-side models \cite{lee2019device}. Since the principle of on-device machine learning is well-aligned with the need for low-cost and privacy-preserving recommender systems, there has been a growing trend towards on-device recommendation \cite{han2021deeprec,wang2020next,ochiai2019real,changmai2019device,chen2021learning,yin2021tt}. However, due to the limited memory and computing power, running a cloud-side recommendation model overloads the edge devices \cite{dhar2021survey}. We have to compress the model to fit resource-constrained environments. Therefore, it is non-trivial to develop practicable on-device recommender systems, which are confronted with two fundamental challenges. 
\par
The first challenge is \textbf{how to reduce the size of regular models to adapt to edge devices}? Unlike the vision and language models \cite{han2015deep} which are usually over-parameterized with a very deep network structure, the recommendation model only contains several intermediate layers that account for a small portion of learnable parameters. Instead, since the recommender systems need to uniquely identify millions of items and users, the embedding table is the one that accounts for the vast majority of memory use \cite{wu2020saec}. Classical model compression techniques such as pruning \cite{srinivas2015data}, down-sizing \cite{hinton2015distilling}, and parameter sharing \cite{plummer2020shapeshifter} hence are less suited, leading to negligible reduced sizes. Previous practices of on-device recommendation \cite{wang2020next,sun2020generic} mostly adopt decomposition techniques such as tensor-train decomposition (TTD) \cite{oseledets2011tensor} to approximate the embedding table with a series of matrix multiplication. However, to avoid drastic performance drop, they only compress the embedding table with a small compression rate (e.g., 4$\sim$8x) and hesitate to explore more compact tensor approximations.
\par
The second challenge is \textbf{how to retain the original capacity}? Inevitably, after being compressed, the capacity of the lightweight model would seriously degrade. To fix this problem, a typical approach is knowledge distillation \cite{hinton2015distilling} whose pipeline is first training a \textit{teacher} model on the cloud by fully utilizing the abundant resources and then transferring teacher's knowledge to the lightweight model, which is called \textit{student}. However, although the existing KD frameworks have been proved beneficial in traditional machine learning problems such as classification and regression, applying it to recommendation is still challenging due to the data sparsity issue \cite{kang2020rrd,lee2019collaborative}. Besides, recent studies find that models with similar structures (e.g., encoder–decoder) are easier to transfer knowledge \cite{chen2021knowledge}, while the tensor decomposition may widen the structural gap between the teacher and the student because it can be seen as a special linear layer prior to student's embedding layer.
\par
In this paper, we work towards an ultra-compact and effective on-device recommendation model under a self-supervised knowledge distillation framework. To tackle the first challenge, given the dilemma that the small TT-rank in tensor-train decomposition leads to under-expressive embedding approximations whereas the larger TT-rank sacrifices the model efficiency \cite{tjandra2017compressing}, we introduce the semi-tensor product (STP) operation \cite{cheng2007survey} to tensor-train decomposition for the extreme compression of the embedding table. This operation endows tensor-train decomposition with the flexibility to perform multiplication between tensors with inconsistent ranks, taking into account both effectiveness and efficiency. Meanwhile, to tackle the second challenge, inspired by the genetic recombination \cite{meselson1975general} in biology, we propose to recombine the embeddings learned by the teacher and the student to perform self-supervised tasks, which can help distill the essential information lying in the teacher's embeddings, so as to compensate for the accuracy loss caused by the extremely high compression. Particularly, we exchange the long-tail item embeddings learned by the two models to form new embeddings. For those long-tail items with few interactions, this \textit{embedding recombination} strategy provides them with the opportunities to learn with the direct guidance from the teacher, reducing the impact of the asymmetric teacher-student structure and leading to more expressive embeddings.
\par
To summarize, our contributions are as follows:
\begin{itemize}[leftmargin=*]
	\item We explore ultra-compact on-device recommendation models by introducing semi-tensor product to tensor-train decomposition for higher compression rates of the embedding table.
	\item We develop a self-supervised KD framework to compensate for the accuracy loss caused by the high compression rate. For the first time, self-supervised learning is applied to on-device recommendation.
	\item We validate the effectiveness of the proposed on-device recommendation model in the next-item recommendation scenario. The extensive experiments on two benchmarks demonstrate that, with 30x model size reduction, the compressed model almost comes with no accuracy loss, and even outperforms its uncompressed counterpart in some cases. The code is released at \url{https://github.com/xiaxin1998/OD-Rec}.
\end{itemize}

\begin{figure*}[t]
	\centering
	\includegraphics[width=0.85\textwidth]{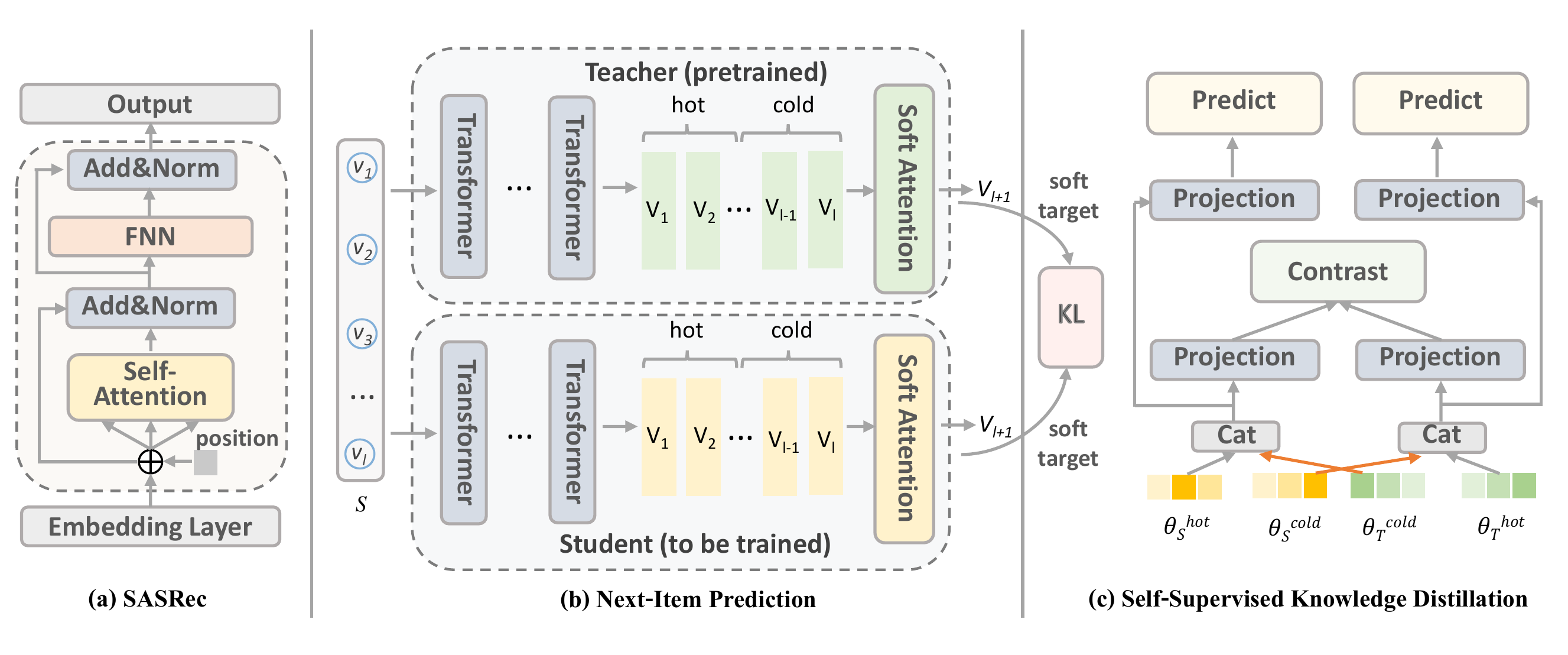}
	\caption{An overview of the proposed framework.}
	\label{figure.1}
\end{figure*}

\section{Related Work}
\subsection{On-Device Recommendation}

On-device machine learning is becoming a trend due to its low latency inference and advantages on privacy protection. This line of research \cite{chen2021learning, changmai2019device, dhar2021survey, han2021deeprec, ochiai2019real, chen2021learning} mainly adopts model compressing techniques like pruning \cite{han2015deep, srinivas2015data}, quantization \cite{gong2014compressing}, and low-rank factorization \cite{novikov2015tensorizing, oseledets2011tensor} to compress regular models so as to fit edge devices. Recently, some on-device recommender models emerged, achieving desired performance with a tiny model size. Chen \textit{et al.} \cite{chen2021learning} proposed to learn elastic item embeddings to make lightweight models adaptively fit various devices under any memory constraints for on-device recommendation. DeepRec \cite{han2021deeprec} uses model pruning and embedding sparsification techniques to fulfil model training on mobile devices and fine-tunes the model using local user data, reaching comparable performance in sequential recommendation with 10x parameter reduction. Wang \textit{et al.} \cite{wang2020next} proposed LLRec, a lightweight next POI recommendation model based on tensor-train factorization. TT-Rec \cite{yin2021tt} applies tensor-train decomposition in the embedding layer to compress the model and improves performance with a sampled Gaussian distribution for the weight initialization of the tensor cores. 

\subsection{Knowledge Distillation for Recommendation}
Knowledge Distillation (KD) \cite{hinton2015distilling} is a typical approach to transfer knowledge from a well-trained teacher model to a simple student model. In this framework, the student model is optimized towards two objectives: minimizing the difference between the prediction and the ground-truth, and fitting the teacher's label distribution or intermediate layer embeddings. KD was first introduced to the classification problem, and currently some KD methods have been proposed for recommender systems \cite{tang2018ranking, lee2019collaborative, kang2020rrd}. The first work is Ranking Distillation \cite{tang2018ranking}, which chooses top-K items from teacher's recommendation list as the external knowledge to guide the student model to assign higher scores when generating recommendations. However, this ignores rich information beyond the recommended top-K items. The follow-up work, CD \cite{lee2019collaborative} proposes a rank-aware sampling method to sample items from the teacher model as knowledge for the student model, enlarging the range of information to be transferred. DE-RRD \cite{kang2020rrd} develops an expert selection strategy and relaxes the ranking-based sampling method to transfer knowledge. Though effective, these KD methods can hardly tackle the cold-start problem caused by the long-tail distribution in recommender systems. There also have been some works \cite{tian2019contrastive,lee2018self} that combine self-supervised learning and knowledge distillation. But they are mainly designed for vision tasks and and cannot be easily tailored for recommendation. 

\subsection{Session-Based Next-Item Recommendation}
The early session-based recommendation \cite{shani2005mdp,rendle2010factorizing, yin2016spatio} is based on Markov Chain and mainly focus on modeling temporal order between items. In the deep learning era, networks such as RNNs \cite{hidasi2015session, chen2020sequence,zhang2018discrete} are modified to model session data and handle the temporal shifts within sessions. The follow-up models, NARM \cite{li2017neural} and STAMP \cite{liu2018stamp} employ attention mechanism to give items different priorities when profiling user's main interests. Graph-based methods \cite{wu2019session, qiuexploiting} construct various types of session graphs to model item relations. For example, SR-GNN \cite{wu2019session} constructs session graphs for every session and designs a gated graph neural network to aggregate information between items into session representations. GCE-GNN \cite{wang2020global} proposes to capture both global-level and session-level interactions and aggregates item information through graph convolution and self-attention mechanism. Xia \textit{et al.} \cite{xia2021aaai, xia2021cikm} proposed to integrate self-supervised learning into session-based recommendation to boost recommendation performance. These models are of great capacity in generating accurate recommendations but they all rely on sufficient storage and computing resources in the cloud, which cannot run on edge devices like smartphones.

\section{Preliminaries}
\subsection{Next-Item Recommendation Task}
In this paper, we apply our on-device recommendation model to the session-based scenario for next-item recommendation. Let $\mathcal{V} = \{v_{1}, v_{2}, v_{3}, ... , v_{|\mathcal{V}|}\}$ denote item set and $s = [v_{s,1}, v_{s,2}, ..., v_{s,l}]$ denote a session sequence. Every session sequence is composed of interacted items in the chronological order from an anonymous user. The task of session-based recommendation is to predict the next item, namely $v_{s,l+1}$, for the current session. In this scenario, every item $v\in \mathcal{V}$ is first mapped into the embedding space. The item embedding table is with a huge size and is denoted as $\mathbf{X}\in\mathbb{R}^{|\mathcal{V}| \times N}$. Given $\mathcal{V}$ and $s$, the output of session-based recommendation model is a ranked list $y = [y_{1}, y_{2}, y_{3}, ..., y_{|\mathcal{V}|}]$ where $y_{v}\ (1 \leq v \leq |\mathcal{V}|)$ is the corresponding predicted probability of item $v$. The top-\textit{K} items $(1 \leq K \leq |\mathcal{V}|)$ with highest probabilities in $y$ will be selected as the recommendations.


\subsection{Tensor-Train Decomposition}
Tensor-train decomposition (TTD) is a typical model compressing technique that can compress a large tensor in a tensor-train (TT) format \cite{oseledets2011tensor} where each element in the tensor can be computed by a sequence of smaller tensor multiplication.
Formally, given a $d$-dimensional tensor $\mathcal{A} \in \mathbb{R}^{N_{1} \times N_{2} \times ... \times N_{d}}$, each entry indexed by ($i_{1}, i_{2}, ..., i_{d}$) can be represented in the following TT-format (for conciseness, we use $x$ to refer to $\mathcal{A}\left(i_{1}, i_{2},..., i_{d}\right)$:
\begin{equation}
\begin{aligned} 
x&=\sum_{r_{1}=1}^{R_{1}} \sum_{r_{2}=1}^{R_{2}}... \sum_{r_{d-1}=1}^{R_{d-1}} \mathbf{G}_{1}\left(i_{1}, r_{1}\right) \mathbf{G}_{2}\left(r_{1}, i_{2}, r_{2}\right)... \mathbf{G}_{d}\left(r_{d-1}, i_{d}\right) \\
&=\underbrace{\mathbf{G}_{1}\left[i_{1},:\right]}_{1 \times R_{1}} \underbrace{\mathbf{G}_{2}\left[:, i_{2},:\right]}_{R_{1} \times R_{2}}...\underbrace{\mathbf{G}_{d-1}\left[:, i_{d-1},:\right]}_{R_{d-2} \times R_{d-1}} \underbrace{\mathbf{G}_{d}\left[:, i_{d}\right]}_{R_{d-1} \times 1}.
\end{aligned}
\end{equation}
$\{\mathbf{G}_{k}\}_{k=1}^{d}$ are called TT-cores and $\mathbf{G}_{k}$ is of the size $R_{k-1} \times N_{k}\times R_{k}$, where $R_{k} \in \left[1, N_{k}\right]$, and $R_{0} = R_{d} = 1$. The sequence of $\{R_{k}\}_{k=0}^{d}$ is called TT-rank. $\mathbf{G}_{k}\left[:, i_{k},:\right]$ represents a slice of tensor $\mathbf{G}_{k}$.\par
If $N_k$ can be further factorized into $I_k\times J_k$, then $\mathbf{G}_{k} \in \mathbb{R}^{R_{k-1} \times N_{k} \times R_{k}}$ can be represented as $\mathbf{G}_{k}^{*} \in \mathbb{R}^{R_{k-1} \times I_k \times J_k \times R_{k}}$. Eq.(1) hence can be rewritten as:
\begin{equation}
\begin{aligned} &
\mathcal{A}\left(\left(i_1, j_1\right), \cdots, \left(i_d, j_d\right)\right) = \mathbf{G}_{1}^{*}\left(i_1, j_1\right) \mathbf{G}_{2}^{*}\left(i_2, j_2\right) \cdots \mathbf{G}_{d}^{*}\left(i_d, j_d\right) \\ &
= \mathbf{G}_{1}^{*}\left[\left(i_1, j_1\right), :\right] \mathbf{G}_{2}^{*}\left[:, \left(i_2, j_2\right), :\right] \cdots \mathbf{G}_{d}^{*}\left[\left(i_d, j_d\right), :\right],
\end{aligned}
\end{equation}
where $0 \leq i_{k}<I_{k}, 0 \leq j_{k}<J_{k}$ ($i_k, j_k$ are integers), $\forall k=1, \ldots, d$. 

\subsection{Semi-Tensor Product}
Semi-tensor product (STP) is a generalization of conventional matrix product \cite{cheng2007survey} and the factor matrices can have arbitrary dimensions. In this paper, we focus on the case when the number of columns in the left matrix is proportional to the number of rows in the right matrix, which is called left semi-tensor product. Let $\mathbf{a} \in \mathbb{R}^{1 \times np}$ denotes a row vector and $\mathbf{b} \in \mathbb{R}^{p}$, then $\mathbf{a}$ can be split into $p$ equal-size blocks as $\mathbf{a}^{1}$, $\mathbf{a}^{2}$, ..., $\mathbf{a}^{p}$, the left semi-tensor product denoted by $\ltimes$ can be defined as:
\begin{equation}
\left\{\begin{array}{l}\mathbf{a} \ltimes \mathbf{b}=\Sigma_{i=1}^{p} \mathbf{a}^{i} \mathbf{b}_{i} \in \mathbb{R}^{1 \times n} \\ \mathbf{b}^{\mathrm{T}} \ltimes \mathbf{a}^{\mathrm{T}}=\Sigma_{i=1}^{p} \mathbf{b}_{i}\left(\mathbf{a}^{i}\right)^{\mathrm{T}} \in \mathbb{R}^{n}.\end{array}\right.
\end{equation}
Then, for two matrices $\mathbf{A} \in \mathbb{R}^{H \times nP}$, $\mathbf{B} \in \mathbb{R}^{P \times Q}$, STP is defined as:
\begin{equation}
\mathbf{C}=\mathbf{A} \ltimes \mathbf{B},
\end{equation}
and $\mathbf{C}$ consists of $H \times Q$ blocks and each block $\mathbf{C}^{hq} \in \mathbb{R}^{1 \times n}$ can be calculated as:
\begin{equation}
\mathbf{C}^{h q}=\mathbf{A}(h,:) \ltimes \mathbf{B}(:, q),
\end{equation}
where $\mathbf{A}(h,:)$ represents $h$-th row in $\mathbf{A}$, $\mathbf{B}(:, q)$ represents $q$-th column in $\mathbf{B}$ and $h$ = 1, 2, $\cdots$, $H$, $q$ = 1, 2, $\cdots$, $Q$. 

\section{Ultra-Compact On-Device Recommendation Model}
\label{st:model}
\subsection{Basic Model Structure}
In this paper, we use SASRec \cite{kang2018self} as our base model. SASRec is a famous Transformer-based \cite{vaswani2017attention} sequential recommendation model in which several self-attention blocks are stacked, including the embedding layer, the self-attention layer and the feed-forward network layers (shown in Fig~\ref{figure.1}.(a)). The embedding layer adds position embeddings to the original item embeddings to indicate the temporal and positional information in a sequence. The self-attention mechanism can model the item correlation. Inside the block, residual connections, the neuron dropout, and the layer normalization are sequentially used.
The model can be formulated as:
\begin{equation}
\mathbf{\hat{X}} = \mathbf{X} + \mathbf{P},\ \mathbf{F} = \mathrm{Attention}(\mathbf{\hat{X}}),\ \mathbf{\Theta} = \mathrm{FNN}(\mathbf{F}),
\end{equation}
where $\mathbf{X}$, $\mathbf{P}$ are the item embeddings and position embeddings, respectively, $\mathbf{F}$ is the session representation learned via self-attention blocks. $\mathrm{FNN}$ represents feed forward network layers and $\mathbf{\Theta}$ aggregates embeddings of all items. In the origin SASRec, the embedding of the last clicked item in $\mathbf{\Theta}$ is chosen as the session representation. However, we think each contained item would contribute information to the session representation for a comprehensive and more accurate understanding of user interests. Instead, we slightly modify the original structure by adopting the soft-attention mechanism \cite{wang2020global} to generate the session representation, which is computed as:
\begin{equation}
	\alpha_{t}=\mathbf{f}^{\top} \sigma\left(\mathbf{W}_{1} \mathbf{x}^{*}_{s}+\mathbf{W}_{2} \mathbf{x}_{t}+\mathbf{c}\right),\,\,\,
	\mathbf{\theta}_{s}=\sum_{t=1}^{l} \alpha_{t} \mathbf{x}_{t},
\end{equation}
where $\mathbf{W}_{1}\in\mathbb{R}^{N \times N}$, $\mathbf{W}_{2} \in\mathbb{R}^{N \times N}$, $\mathbf{c}, \mathbf{f}\in\mathbb{R}^{N}$ are learnable parameters, $\mathbf{x}_{t}$ is the embedding of item $t$ and $\mathbf{x}_{s}^{*}$ is obtained by averaging the embeddings of items within the session $s$, i.e. $\mathbf{x}_{s}^{*} = \frac{1}{l}\sum_{t=1}^{l}\mathbf{x}_{t}$. $\sigma$ is sigmoid function. Session representation $\mathbf{\theta}_{s}$ is represented by aggregating item embeddings while considering their corresponding importances. 

\begin{figure}[t]
	\centering
	\includegraphics[width=0.4\textwidth]{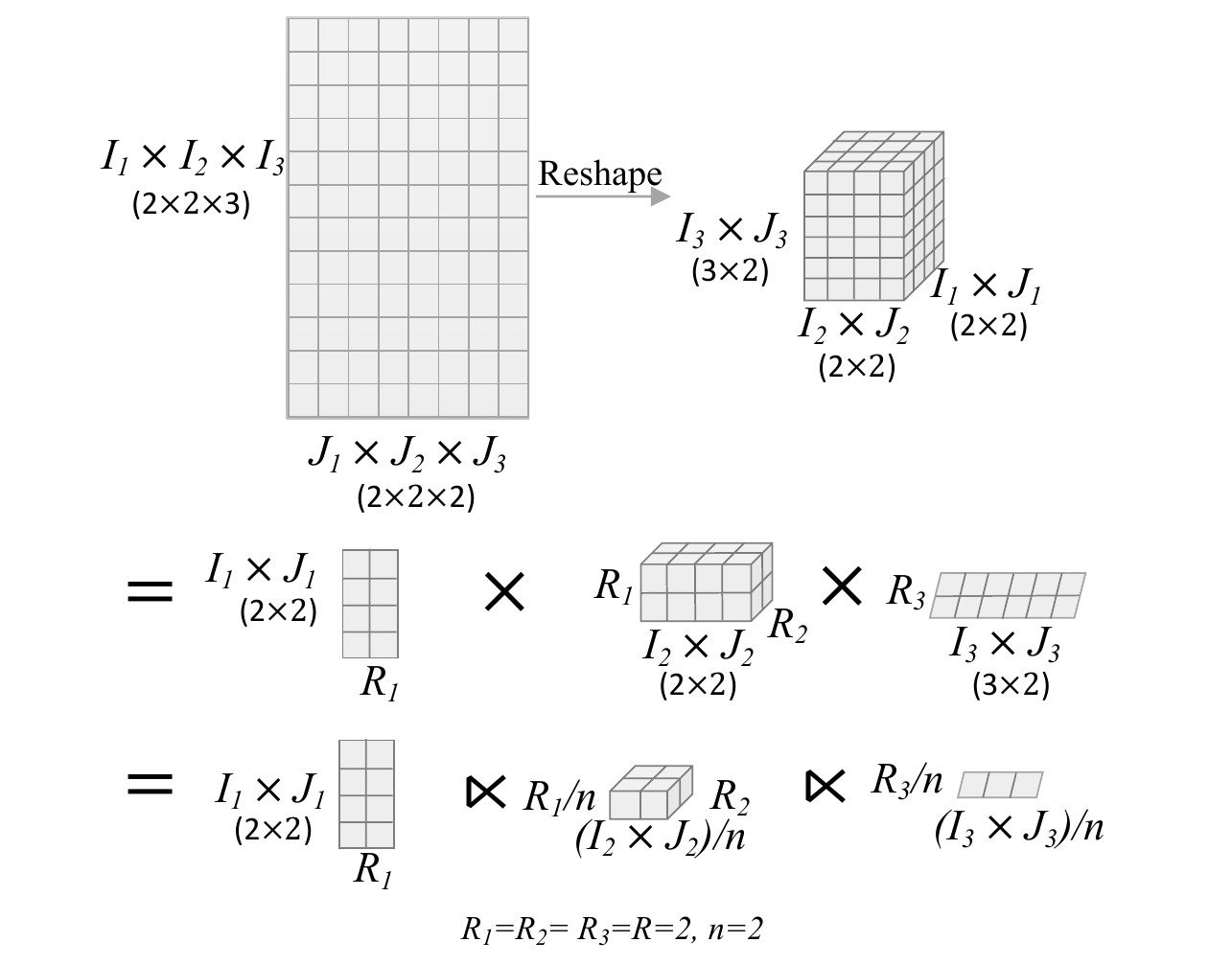}
	\caption{Given the embedding table of size 12$\times$ 8, after applying TTD, it can be approximated by the multiplication of three small tensors. While after applying STTD, the number of the parameters is further compressed.}
	\label{figure.2}
	\vspace{-10pt}
\end{figure}
\subsection{Model Compression}
It is obvious that the item embedding table accounts for the large majority of total learnable parameters. Compressing the embedding table is the key to employ on-device recommendation. Therefore, we first adopt the tensor-train decomposition (TTD) \cite{hrinchuk2019tensorized, novikov2015tensorizing} to shrink the size of item embeddings. 
To factorize the embedding table into smaller tensors, we let the total number of items $|V| = \prod_{k=1}^{d} I_{k}$ and the embedding dimension $N = \prod_{k=1}^{d} J_{k}$. For one specific item $v$ indexed by $i$, we map the row index $i$ into $d$-dimensional vectors $\mathbf{i}$ = $\left(i_1, ..., i_d\right)$ according to \cite{hrinchuk2019tensorized}, and then get the particular slices of the index in TT-cores to perform matrix multiplication on the slices (Eq. (1)). Benefitting from this decomposition, the model does not need to store the embedding table which consumes large memory. Instead we just store the TT-cores to fulfil an approximation of the entry in the embedding table, i.e., a sequence of small tensor multiplication. Given $\{\mathbf{G}_{k}\}_{k=1}^{d}$ of size $\sum_{k=1}^{d} R_{k-1} I_{k} J_{k} R_{k}$, the compression rate is:
\begin{equation}
\text { rate }=\frac{\prod_{k=1}^{d} I_{k} J_{k}}{\sum_{k=1}^{d} R_{k-1} I_{k} J_{k} R_{k}} = 
\frac{\prod_{k=1}^{d} I_{k} J_{k}}{I_1 J_1 R + \sum_{k=2}^{d-1} I_{k} J_{k} R^2 + I_d J_d R},
\end{equation}
where $R_0=R_d=1$. For simplicity, we assign the same rank $R$ to the intermediate TT-cores, namely, $\{R_k\}_{k=2}^{d-1} = R$ in this paper. \par
However, conventional tensor-train decomposition requires strict dimensionality consistency between factors and probably leads to dimension redundancy. For example, if we decompose the item embedding matrix into two smaller tensors, i.e., $\mathbf{X} = \mathbf{G}_{1}\mathbf{G}_{2}$, to get each item's embedding, the number of columns of $\mathbf{G}_{1}$ should be the same with the number of rows of $\mathbf{G}_{2}$. This consistency constraint is too strict for flexible and efficient tensor decomposition, posing an obstacle to further compression of the embedding table. Therefore, inspired by \cite{zhao2021semi}, we propose to integrate semi-tensor product with tensor-train decomposition where the dimension of TT-cores can be arbitrarily adjusted. The conventional tensor product between TT-cores is replaced with the semi-tensor product as:
\begin{equation}
\mathbf{X} = \hat{\mathbf{G}}_1 \ltimes \hat{\mathbf{G}}_2 \ltimes \cdots \ltimes \hat{\mathbf{G}}_d,
\end{equation}
where $\{\hat{\mathbf{G}}_{k}\}_{k=1}^{d}$ are the core tensors after applying semi-tensor product based tensor-train decomposition (STTD) and $\hat{\mathbf{G}}_{1} \in \mathbb{R}^{I_1 J_1 \times R}$, $\{\hat{\mathbf{G}}_{k}\}_{k=2}^{d-1} \in \mathbb{R}^{ \frac{R}{n} \times \frac{I_{k} J_k}{n} \times R}, \hat{\mathbf{G}}_{d} \in \mathbb{R}^{\frac{R}{n} \times \frac{I_d J_d}{n}}$.
The compression rate is:
\begin{equation}
\text { rate }=\frac{\prod_{k=1}^{d} I_{k} J_{k}}{I_1 J_1 R + \sum_{k=1}^{d-1} I_{k} J_{k} \frac{R^2}{n^2} + I_d J_d \frac{R}{n^2}}.
\end{equation}
We can adjust the TT-rank, namely, the values of the hyperparameters $R$ and $n$, and the length $k$ of the tensor chain to flexibly compress the model in any large degree. In Fig.2 we illustrate TTD and STTD to facilitate understanding. 
\par
Putting Eq. (8) and Eq. (10) together, we can easily find that when $k$ is large enough, the compression rate calculated by Eq. (10) approaches $n^{2}$ times the rate derived from Eq. (8), towards an ultra-compact model. We hereby show the model compression ability with different TT-ranks by the example in Table 1. We assume there are 20,000 items and this number is factorized into $10 \times 10 \times 25 \times 8$. The original item embedding dimension is 128 which is factorized into $4 \times 4\times4\times2$. We calculate the model size before and after applying TTD and STTD with different TT-ranks and $n$ = 2. As shown in the table, it is clear that, with the TT-rank getting smaller, the number of the parameters is dramatically compressed by several orders of magnitude. Particularly, with the same TT-rank, STTD further shrinks the model to about 1/3$\sim$1/4 the size compressed by TTD. Meanwhile, our experiments in section 6 demonstrate that, the further compression supported by STTD will not incur performance drop on the one compressed by TTD.

\begin{table}[h]
    \renewcommand\arraystretch{1.0}
    \label{Table:1}
    \begin{center}
        \begin{tabular}{c|c|cc|cc}
            \toprule
            \multirow{2}{*}{Original Size} & \multirow{2}{*}{TT-rank} & \multicolumn{2}{c}{TTD} & \multicolumn{2}{c}{STTD}\cr
            & & size & rate & size & rate\\
            \hline
             & 4 & 2464 & 1039 & 736 & 3478   \\
            2,560,000 & 8 & 9408& 272 & 2592 &988  \\
             & 16 & 36736 & 70 &9664 & 265 \\
             \bottomrule
        \end{tabular}
    \end{center}
    \caption{Model Size Analytics with Different TT-ranks in TTD and STTD where $n$ = 2.}
	\vspace{-20pt}
\end{table}

\section{Self-Supervised Knowledge Distillation}
A highly compressed model brings efficiency but inevitably comes at the cost of capacity loss compared with the original. Therefore, in this section we propose a novel self-supervised knowledge distillation framework to restore the capacity. In this framework, the compressed model plays the role of student and is trained with the help of its pre-trained uncompressed counterpart, namely, the teacher. Note that the only difference between the teacher and the student model is that the embedding table of the student is compressed by STTD. \par

\subsection{Architecture}
To the best of our knowledge, we are the first to apply self-supervised knowledge distillation to recommendation. Therefore, we first define a general framework of self-supervised knowledge distillation in the scenario of recommendation. Let $\mathcal{D}$ denote the item interaction data, $\mathcal{D}_{soft}$ represent the soft targets from the teacher and $\mathcal{D}_{ssl}$ denote the augmented data in the self-supervised setting. The teacher model and the student model are denoted by $M_{T}$ and $M_{S}$, respectively. Then the framework is formulated as follows:
\begin{equation}
f\left(M_{T}, M_{S}, \mathcal{D}, \mathcal{D}_{soft}, \mathcal{D}_{ssl} \right) \rightarrow M_{S}^{*},
\end{equation}
which means that by jointly supervising the student model with the historic interactions $\mathcal{D}$, the teacher's soft targets $\mathcal{D}_{soft}$ and the self-supervised signals extracted from $\mathcal{D}_{ssl}$, we can finally obtain an improved student model $M_{S}^{*}$ that retains the capacity of the regular model. We define the learning objective of the student as:
\begin{equation}
\mathcal{L} = \alpha_{r}\mathcal{L}_{rec} + \vec{\mathbf{\beta}}^{\top}\mathcal{L}_{kd}, \,\,\,
\mathcal{L}_{kd} = \{\mathcal{L}_{kd}^{1}, \mathcal{L}_{kd}^{2},...\}.
\end{equation}
$\mathcal{L}_{rec}$ is the recommendation loss, $\mathcal{L}_{kd}$ is the KD loss which is composed of multiple tasks. $\alpha_{r}$ and $\vec{\mathbf{\beta}}$ are hyperparameters to control the magnitudes of different tasks. Architecturally, the proposed framework is model-agnostic so as to boost recommendation models with arbitrary structures. Note that, the teacher model has been well-trained before and will not be updated in this framework. We design three types of distillation tasks: traditional distillation task based on soft targets, contrastive self-supervised distillation task and predictive self-supervised distillation task.

\subsection{Embedding Recombination}
The essential idea of self-supervised learning \cite{liu2020self,yu2021socially,yu2021www} is to learn with the supervisory signals which are extracted from augmentations of the raw data. However, in recommender systems, interactions follow a long-tail distribution \cite{yu2018adaptive}. For the long-tail items with few interactions, generating informative augmentations is often intractable. Inspired by the genetic recombination \cite{meselson1975general} and the preference editing in \cite{ma2021improving}, we come up with the idea to exchange embedding segments from the student and the teacher and recombine them to distill knowledge. Such recombinations fulfil the direct information transfer between the teacher and the student, and meanwhile create representation-level augmentations which inherit characteristics from both sides. Under the direct guidance from the teacher, we expect that the student model can learn more from distillation tasks. Since the parameters of the teacher are frozen, this strategy will not impact the teacher model. We term this method \textit{embedding recombination}. 
\par
To implement it, we divide items into two types: hot and cold by their popularity. The top 20\% popular items are considered as hot items and the rest part is cold items. For each session, we split it into two sub-sessions: cold session and hot session in which only the cold items or hot items are contained. Then we learn two corresponding session representations separately: hot session representation and cold session representation. The hot session representation derives from representations of hot items in that session based on the soft attention mechanism in Eq. (7), while the cold session representation is analogously learned from representations of cold items. They are formulated as follows:
\begin{equation}
\mathbf{\theta}_{s}^{hot}=\sum_{t=1}^{l_{h}} \alpha_{t}^{hot} \mathbf{x}_{t}^{*hot}, \,\,\,\mathbf{\theta}_{s}^{cold}=\sum_{t=1}^{l_{c}} \alpha_{t}^{cold} \mathbf{x}_{t}^{*cold},
\end{equation}
where $l_{c}$ and $l_{h}$ are lengths of the cold session and the hot session, $\mathbf{x}_{t}^{*hot}(\mathbf{x}_{t}^{*cold})$ represents representation of the t-$th$ hot item or cold item in the session $s$, $\alpha_{t}^{hot}(\alpha_{t}^{cold})$ is the learned attention coefficient of the t-$th$ hot or cold item in the session, and $\mathbf{\theta}^{hot}(\mathbf{\theta}^{cold})$ is the learned hot or cold session representation. In both the student and the teacher models, we learn such session embeddings. Then for the same session, we swap their cold session representations as follows to generate new embeddings:
\begin{equation}
z_{s}^{tea} =[\mathbf{\theta}_{s}^{hot_{tea}}, \mathbf{\theta}_{s}^{cold_{stu}}],\,\,\, z_{s}^{stu} = [\mathbf{\theta}_{s}^{hot_{stu}}, \mathbf{\theta}_{s}^{cold_{tea}}].
\end{equation}
As for those sessions which only contain hot items or cold items, we generate the corresponding type of session representations and swap them. The embedding recombination is illustrated in Fig. 1.(c).

\subsection{Contrastive Self-Supervised Task}
Since the generated new embeddings are profiling the same session, we consider that there is shared learnable invariance lying in these embeddings. Therefore, we propose to contrast recombined embeddings so as to learn discriminative TT-core representations for the student model, which can mitigate the data sparsity issue to some degree. Meanwhile, this contrastive task also helps to transfer the information in the teacher's representation, rather than only transferring the output probabilities. We follow \cite{yu2022graph} to use InfoNCE \cite{oord2018representation} to maximize the mutual information between teacher's and student's session representations. For any session $s$, its recombined representations are positive samples of each other, while the recombined representations of other sessions are its negative samples. We conduct negative sampling in the current batch $\mathcal{B}$. The loss of the contrastive task is defined as follows:
\begin{equation}
\mathcal{L}_{cl}=-\underset{s\in\mathcal{B}}{\sum}\log \frac{ \psi\left(\mathbf{W}_{t} z_{s}^{tea}, \mathbf{W}_{s} z_{s}^{stu}\right)}{ \psi\left(\mathbf{W}_{t} z_{s}^{tea}, \mathbf{W}_{s} z_{s}^{stu}\right)+\underset{j\in\mathcal{B}/\{s\} }{\sum} \psi\left(\mathbf{W}_{t} z_{s}^{tea}, \mathbf{W}_{s} z_{j}^{stu}\right)},
\end{equation}
where $\psi\left(\mathbf{W}_{t} z_{s}^{tea}, \mathbf{W}_{s} z_{s}^{stu}\right)$ = $\exp\left(\phi\left(\mathbf{W}_{t} z_{s}^{tea}, \mathbf{W}_{s} z_{s}^{stu}\right) / \tau\right)$, $\phi\left(\cdot\right)$ is the cosine function, $\tau$ is temperature (0.2 in our method), and $\mathbf{W_{t}}$ and $\mathbf{W_{s}}\in \mathbb{R}^{N \times 2N}$ are projection matrices.  \par

\subsection{Predictive Self-Supervised Task}
The contrastive task directly distills the recombined embeddings whilst ignoring the ground truth. We consider that under the direct guidance of the session embedding from the teacher model, the swapped session embedding from the student can improve its ability to predict the ground truth. We then let the teacher and student learn from the ground-truth using their recombined session representations, which also makes the KD framework compatible with the main recommendation task. The predictive loss is defined as follows:
\begin{equation}
\begin{split}
&
\mathcal{L}_{pred}=-\left(\sum_{v=1}^{|\mathcal{V}|} \mathbf{y}_{v} \log \left(\hat{\mathbf{y}}_{v}^{tea}\right)+\left(1-\mathbf{y}_{v}\right) \log \left(1-\hat{\mathbf{y}}_{v}^{tea}\right)\right)\\
& -
\left(\sum_{v=1}^{|\mathcal{V}|} \mathbf{y}_{v} \log \left(\hat{\mathbf{y}}_{v}^{stu}\right)+\left(1-\mathbf{y}_{v}\right) \log \left(1-\hat{\mathbf{y}}_{v}^{stu}\right)\right),
\end{split}
\end{equation}
where $\mathbf{y}$ is one-hot encoding vector of ground-truth, $\hat{\mathbf{y}}_{v}^{tea}$ represents the predicted score of the teacher on item $v$ for each session $s$, $\hat{\mathbf{y}}^{tea}=\operatorname{softmax}(\left(\mathbf{W}_{tea} z_{s}^{tea}\right)^{\top}\mathbf{x}_{v}^{tea})$, and $\mathbf{W}_{tea}\in \mathbb{R}^{N \times 2N}$. $\hat{\mathbf{y}}_{v}^{stu}$ is computed analogously.

\subsection{Distilling Soft Targets}
The self-supervised KD tasks rely on data augmentations. Though effective, there may be divergence between original representations and recombined representations. We finally distill the soft targets (i.e., teacher's prediction probabilities on items). Following the convention, we adopt KL Divergence to make the student generate probabilities similar to the teacher's. Let $\mathrm{prob}^{tea}$ and $\mathrm{prob}^{stu}$ represent the predicted probability distributions of the teacher and the student, respectively. Then for each session $s$, we have:
\begin{equation}
\begin{aligned}
\mathrm{prob}^{tea} = \mathrm{softmax}(\mathbf{\theta}_{s}^{tea\top}\mathbf{X}),\\
\mathrm{prob}^{stu} = \mathrm{softmax}(\mathbf{\theta}_{s}^{stu\top}\mathbf{X});
\end{aligned}
\end{equation}
$\mathbf{\theta}_{s}^{tea}$ and $\mathbf{\theta}_{s}^{stu}$ are representations of session $s$ from teacher and student. Then to maximize the agreement between them, we define the loss of distilling soft targets as:
\begin{equation}
\mathcal{L}_{soft} = - \sum_{v=1}^{|\mathcal{V}|} \mathrm{prob}^{tea}_v \ln \frac{\mathrm{prob}^{stu}_v}{\mathrm{prob}^{tea}_v}.
\end{equation}

\subsection{Model Optimization}
After training using knowledge distillation, we get refined item and session representations. The inner product of one candidate item $v \in \mathcal{V}$ and the given session $s$ is the score of item $v$ to be recommended to this session. We apply softmax to compute the probability:
\begin{equation}
\hat{\mathbf{y}}=\operatorname{softmax}(\mathbf{\theta}_{s}^{\top}\mathbf{x}_{v}).
\end{equation}
We then use the cross entropy loss to be the learning objective:
\begin{equation}
\mathcal{L}_{rec}=-\sum_{v=1}^{|V|} \mathbf{y}_{v} \log \left(\hat{\mathbf{y}}_{v}\right)+\left(1-\mathbf{y}_{v}\right) \log \left(1-\hat{\mathbf{y}}_{v}\right).
\end{equation}
$\mathbf{y}$ is the one-hot encoding vector of the ground truth. For simplicity, we leave out the $L_{2}$ regularization terms.
We jointly learn the recommendation task with the proposed self-supervised knowledge distillation framework through Eq. (12) which is:
\begin{equation}
\mathcal{L} = (1 - \beta_3)\mathcal{L}_{rec} + \beta_1 \mathcal{L}_{cl} + \beta_2 \mathcal{L}_{pred} + \beta_3 \mathcal{L}_{soft}.
\end{equation}
In the recommendation loss, we use a coefficient $\beta_{3}$ to balance the recommendation task and the task of distilling soft targets.

 

\section{Experiments}
\subsection{Experimental Settings}
\subsubsection{Datasets.}
We evaluate our model using two real-world benchmark datasets: \textit{Tmall}\footnote{https://tianchi.aliyun.com/dataset/dataDetail?dataId=42} and \textit{RetailRocket}\footnote{https://www.kaggle.com/retailrocket/ecommerce-dataset}.
\textit{Tmall} is from IJCAI-15 competition and contains anonymized user's shopping logs on Tmall shopping platform. \textit{RetailRocket} is
a dataset on a Kaggle contest published by an E-commerce company, including the user's browsing activities within six months. 
For convenient comparison, we duplicate the experimental environment in \cite{wu2019session, wang2020global}. Specifically, we filter out all sessions whose length is 1 and items appearing less than 5 times. The latest one interaction of each session is set to be the test set and the previous data is used for training. The validation set is randomly sampled from the training set and makes up 10\% of it. Then, we augment and label the training and test datasets by using a sequence splitting method, which generates multiple labeled sequences with the corresponding labels $([v_{s,1}], v_{s,2}), ([v_{s,1},v_{s,2}], v_{s,3}), ...,
([v_{s,1}, v_{s,2}, ..., v_{s,l-1}], v_{s,l})$ for every session $s = [v_{s,1}, v_{s,2}, v_{s,3}, ..., v_{s,l}]$. Note that the label of each sequence is the last consumed item in it. The statistics of datasets are presented in Table 2.

\begin{table}[ht]
	\footnotesize
	\renewcommand\arraystretch{1.0}
	\caption{Dataset Statistics}	
	\label{Table:stat}
	\begin{center}
		\begin{tabular}{l|c|c|c|c}
			\hline
			Dataset&\#Training sessions & \#test sessions &  \#Items & Avg. Length\\ \hline
			\hline			
			Tmall & 351,268 & 25,898 & 40,728 & 6.69\\
			RetailRocket&433,643  &15,132 & 36,968 & 5.43\\
			\hline
		\end{tabular}
	\end{center}
	\vspace{-10pt}
\end{table}

\subsubsection{Baseline Methods.}
We compare our method with the following representative session-based recommendation methods:
\begin{itemize}[leftmargin=*]
	\item \textbf{GRU4Rec} \cite{hidasi2015session} is a GRU-based session-based recommendation model which also utilizes a session-parallel mini-batch training process and adopts ranking-based loss functions to model user sequences.
	\item \textbf{NARM} \cite{li2017neural}: is an RNN-based model which employs attention mechanism to capture user's main purpose and combines it with the temporal information to generate recommendations.
	\item \textbf{STAMP} \cite{liu2018stamp}: adopts attention layers to replace all RNN encoders and employs the self-attention mechanism \cite{vaswani2017attention} to model long-term and short-term user interests.
	\item \textbf{SR-GNN} \cite{wu2019session}:
	proposes a gated graph neural network to refine item embeddings and also employs a soft-attention mechanism to compute the session embeddings.
\end{itemize}
\begin{table}[tp]
		\small
	\label{Table:3}
	\begin{center}
		\setlength{\tabcolsep}{1mm}{
		{
		{
			\begin{tabular}{*{9}{c}}
				\toprule
				\multirow{2}{*}{Method} &
				\multicolumn{4}{c}{Tmall} & \multicolumn{4}{c}{RetailRocket} \cr
				\cmidrule(lr){2-5}\cmidrule(lr){6-9} & P@5 & M@5 & P@10 & M@10 & P@5 & M@5 & P@10 & M@10  \\ \hline		
							
				GRU4REC  & 16.48 & 11.92 & 19.58 & 12.33 & 32.37 & 23.86 & 39.35 & 24.67 \\
				
				NARM & 16.84 & 12.34 & 19.68 &12.72  & 32.94 & 23.36  & 39.18 & 24.19   \\
				
				STAMP  & 17.45 & 12.68  &22.61  & 13.12 & 33.57 & 23.30 & 39.75 & 24.51  \\
				
				SR-GNN & 19.39 & 12.88 & 23.79 & 13.46 & 35.68 & 24.72 & 43.31 & 25.75  \\
				
				Teacher & \textbf{22.47} & \textbf{15.75} & \textbf{27.86} & \textbf{16.32} &\textbf{36.51} &\textbf{24.87}&\textbf{43.59}&\textbf{25.79}\\ \hline
		\end{tabular}}}}		
	\end{center}
	\caption{Performance comparison for all the methods.}
	\vspace{-20pt}
\end{table}

Following \cite{wu2019session}, we use P@K (Precision) and MRR@K (Mean Reciprocal Rank) to evaluate the recommendation results where K is 5 or 10. Precision measures the ratio of hit items while MRR measures the item ranking in the recommendation list. 
\subsubsection{Hyperparameter Settings}
As for the setting about the general hyperparameters, we set the mini-batch size to 100, the $L_2$ regularization to $10^{-5}$, and the embedding dimension to 128 for Tmall and 256 for RetailRocket. All the learnable parameters are initialized with the Uniform Distribution $U(-0.1,0.1)$. And in our base model, the number of attention layer and attention head are 1 and 2 for RetailRocket; 1 and 1 for Tmall. Dropout rate is 0.5 for Tmall and 0.2 for RetialRocket. We use Adam with the learning rate of 0.001 to optimize the model. For all baselines, we report their best performances with the same general experimental settings.

\subsection{Capacity of Teacher Model}
To validate if our base model is qualified to be a good teacher model, we compare it with all the baselines on the two datasets and present results in Table 3. It is clearly shown that our teacher model, which employs Transformer-based structure to learn item representations outperforms other baselines on the two datasets. These results demonstrate that this teacher model is with the adequate capacity that can teach the highly compressed student model.  

\subsection{Effectiveness of Model Compressing}

To evaluate the effectiveness and the generalization ability of the proposed method, we conduct a comprehensive study by investigating different compression rates. There are three factors in the student that can influence the compression rate: the shape of the factorized tensors, TT-rank $R$ and hyperparameter $n$ in semi-tensor product. To investigate the influence of factorization shape, we design four different student models using different tensor shapes, which are denoted as $\textbf{Stu-1}$, $\textbf{Stu-2}$, $\textbf{Stu-3}$ and $\textbf{Stu-4}$. And in those four models, the item embedding table for Tmall is decomposed into tensors with these shapes: \\
$\textbf{Stu-1}: (1, 169\times16, R)\times(R/n, 241\times8/n, 1)$\\
$ \textbf{Stu-2}:(1, 169\times32, R)\times(R/n, 241\times4/n, 1)$\\
$ \textbf{Stu-3}: (1, 13\times8, R)\times(R/n, 13\times4/n, R)\times(R/n, 241\times4/n, 1)$\\
$\textbf{Stu-4}:(1, 13\times8, R)\times(R/n, 13\times8/n, R)\times(R/n, 241\times2/n, 1)$. \\ (The total number of items 40728$\approx$40729=169$\times$241= 13$\times$13$\times$241; the embedding dimension 128=16$\times$8=32$\times$4=8$\times$4$\times$4= 8$\times$8$\times$2). \\
And for RetailRocket, the core tensor shapes of the four models are:\\
$\textbf{Stu-1}:(1, 117\times16, R)\times(R/n, 316\times16/n, 1)$\\
$ \textbf{Stu-2}:(1, 117\times32, R)\times(R/n, 316\times8/n, 1)$\\
$ \textbf{Stu-3}:(1, 18\times8, R)\times(R/n, 26\times8/n, R)\times(R/n, 79\times4/n, 1)$\\
$\textbf{Stu-4}:(1, 18\times16, R)\times(R/n, 26\times4/n, R)\times(R/n, 79\times4/n, 1)$. \\
 (The total number of items 36968$\approx$36972 =117$\times$316 =18$\times$26$\times$79; the embedding dimension 256=16$\times$16=32$\times$8= 8$\times$8$\times$4=16 $\times$4$\times$4). \\
We report the compression rates (CR) and corresponding performances of the above four variants in Table 4 where we use $R$=60, $n$=2 for Tmall and $R$=100, $n$=2 for RetailRocket ($\textbf{Stu-1-60}$ means model $\textbf{Stu-1}$ with $R$=60). It is apparent that the factorization shape has a great influence on the recommendation performance. To our surprise, on the two datasets, by comparing \textbf{Stu-1} with \textbf{Stu-2} and comparing \textbf{Stu-3} with \textbf{Stu-4}, we find that the models which have smaller compression rates do not show better performance on most metrics. Such a finding is quite counter-intuitive. Though the reason is not clear, we suggest that, for a better performance, it is necessary to try different factorization shapes instead of empirically choosing a larger one.\par

\begin{table}[t]
\footnotesize
	\label{Table:4}
	\begin{center}
	\setlength{\tabcolsep}{1mm}{{{
		\begin{tabular}{cccccc|cccccc}
	        \toprule
            \multicolumn{6}{c}{Tmall} & \multicolumn{6}{c}{RetailRocket} \\
			\hline
			\footnotesize{Stu}& \footnotesize{CR} & \footnotesize{P@5} & \footnotesize{M@5} & \footnotesize{P@10} & \footnotesize{M@10} &\footnotesize{Stu}&\footnotesize{CR} & \footnotesize{P@5} & \footnotesize{M@5} & \footnotesize{P@10} & \footnotesize{M@10} \\ \hline
			1-60 & 27 & 23.46 & \textbf{17.79}& \textbf{25.22}&\textbf{18.03} & 1-100 &30 &\textbf{35.17}&\textbf{26.40}&\textbf{36.81}&\textbf{26.63} \\
			2-60 & 15&\textbf{23.68}&17.59&24.87&17.82 & 2-100  &22 &35.03&26.21&36.02&26.37\\
			3-60 &77&21.36&16.52&23.15&16.76 & 3-100  &17 &33.32&25.41&34.23&25.88\\
			4-60 &49&20.77&16.17&22.52&16.41& 4-100  &32 &34.81&25.92&35.57&26.10 \\
			\hline
		\end{tabular}}}}
	\end{center}
	\caption{Performances with different tensor factorizations.}
	\vspace{-10pt}
\end{table}

To explore the relation between TT-rank and recommendation performance, we further experiment with different $R$ values on $\textbf{Stu-1}$ and $\textbf{Stu-3}$ with same $n$ value ($n$=2). For Tmall, we select 20, 40, and 60 as the value of $R$ for $\textbf{Stu-1}$ and 60, 80, 100 for $\textbf{Stu-3}$. For RetailRocket, we choose 40, 60, and 80 for both models. 
We report the results in Table 5. From the table, it is discovered that when using smaller TT-rank in $\textbf{Stu-1}$ or $\textbf{Stu-3}$, the model size decreases and the performance will drop on four measures on the two datasets.\par

\begin{table}[t]
	\footnotesize
		\label{Table:5}
		\begin{center}
		\setlength{\tabcolsep}{1mm}{{{
	\begin{tabular}{cccccc|cccccc}
	\toprule
				\multicolumn{6}{c}{Tmall} & \multicolumn{6}{c}{RetailRocket} \\
				\hline
				 \footnotesize{Stu}& \footnotesize{CR} & \footnotesize{P@5} & \footnotesize{M@5} & \footnotesize{P@10} & \footnotesize{M@10} &\footnotesize{Stu}&\footnotesize{CR} & \footnotesize{P@5} & \footnotesize{M@5} & \footnotesize{P@10} & \footnotesize{M@10} \\ \hline
	
				1-20 &82&21.06&16.47&23.05&16.74 & 1-40&75 &30.15&24.88&31.03&24.66 \\ 
				1-40 & 41&22.69&17.33&24.51&17.58  &1-60 &50 &33.54&25.66&34.88&25.84\\
				1-60 & 27 & 23.46 & 17.79& 25.22&18.03&1-80 &38& 34.97&26.31&36.46&26.52\\
				3-60&77&21.36&16.52&23.15&16.76 &3-40 &102 &29.71&24.22&30.51&24.47\\
				3-80&47&21.61&16.73&23.42&16.97&3-60 &47&33.28&25.26&34.57&25.46\\
				3-100&32&22.49&17.16&24.26&17.40&3-80&26&35.01&26.33&36.52&26.49\\        \hline          
	\end{tabular}}}}
		\end{center}
		\caption{Performances with different TT-ranks.}
		\vspace{-10pt}
	\end{table}

To investigate the effects of hyperparameter $n$ in STTD, we choose different $n$ values for $\textbf{Stu-1}$ with two TT-ranks on each dataset because $\textbf{Stu-1}$ shows the best performance in previous experiments. The TT-rank and $n$ value are denoted in the form of $\textbf{R-n}$. The results shown in Table 6 indicate that, compared with the factorization shape and TT-rank, the performance is more sensitive to the value of $n$. Increasing its value leads to a drastic performance drop. Obviously, there is a trade-off between performance and model size for on-device recommendation.
\begin{table}[t]
	\footnotesize
		\label{Table:6}
		\begin{center}
		\setlength{\tabcolsep}{1mm}{{{
			\begin{tabular}{cccccc|cccccc}
				\toprule
				\multicolumn{6}{c}{Tmall} & \multicolumn{6}{c}{RetailRocket} \\
				\hline
				R-n & CR & P@5 & M@5 & P@10 & M@10 &R-n&CR & P@5 & M@5 & P@10 & M@10 \\ 
				\hline
				40-2&  41&22.69&17.33&24.51&17.58&40-2&75&30.15&24.88&31.03&24.66 \\
				40-4&  46&21.30&16.56&23.35&16.84&40-4&108&28.31&22.01&28.47&22.31  \\
				60-2 & 27&23.46&17.79&25.22&18.03&60-2&50&33.54&25.66&34.88&25.84 \\
				60-4&  31&22.20&17.07&24.17&17.34&60-4&72&30.28&24.79&32.51&24.93 \\				
				\hline
			\end{tabular}}}}
		\end{center}
		\caption{Performances with different values of $n$ in STTD.}
		\vspace{-10pt}
	\end{table}

\begin{figure}[t]
	\centering
	\includegraphics[width=0.48\textwidth]{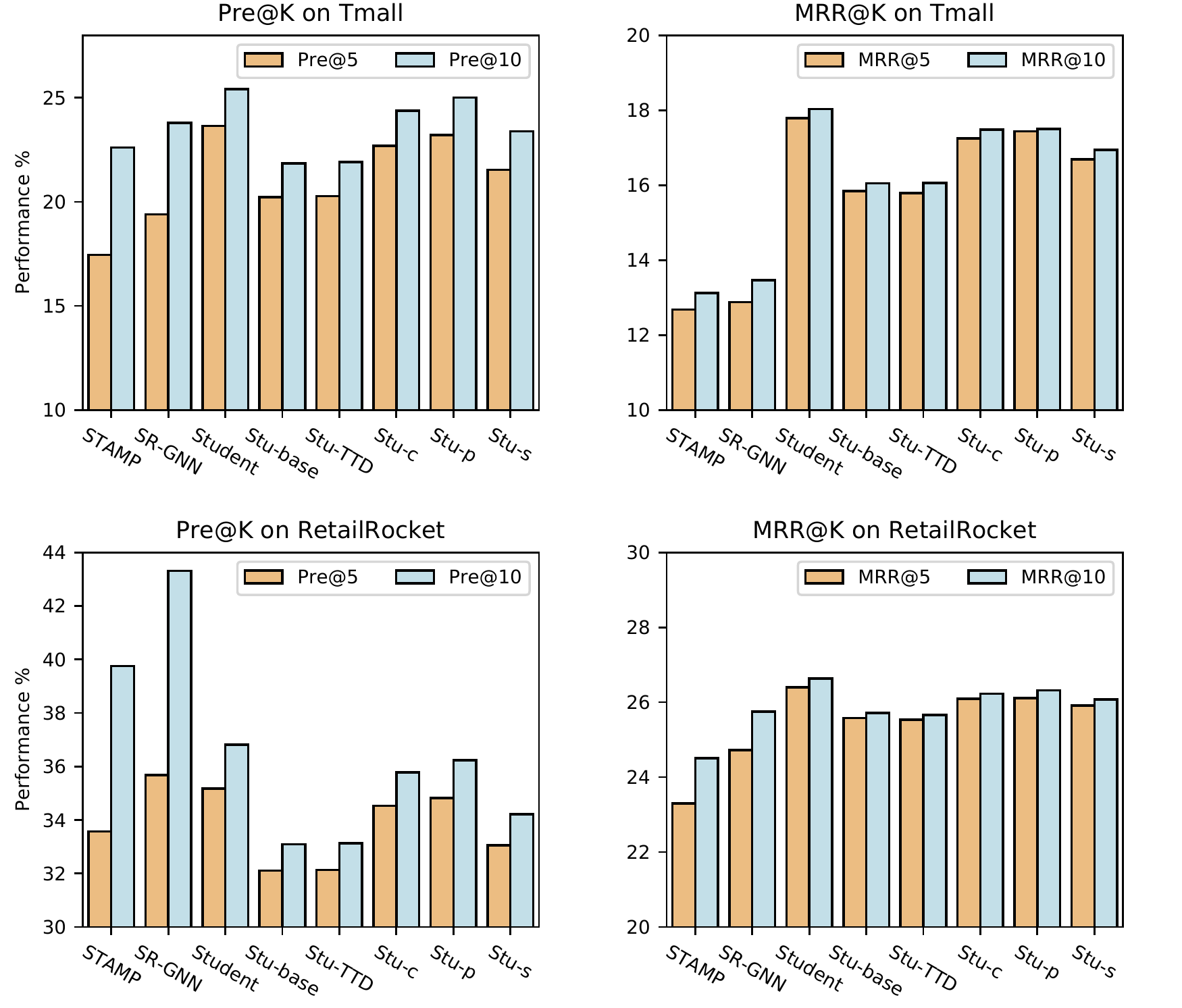}
	\caption{Performances of all the variants of the student.}
	\label{figure.3}
	\vspace{-10pt}
\end{figure}
	
\subsection{Ablation Study}
To investigate the effectiveness of the STTD and the self-supervised knowledge distillation framework, we design five variants: $\textbf{Stu-base}$, $\textbf{Stu-TTD}$, $\textbf{Stu-c}$, $\textbf{Stu-p}$, and $\textbf{Stu-s}$. $\textbf{Stu-base}$ means that the self-supervised knowledge distillation is disabled; $\textbf{Stu-TTD}$ refers to the case where the conventional tensor-train decomposition is used; $\textbf{Stu-c}$, $\textbf{Stu-p}$, and $\textbf{Stu-s}$ correspond to the models in which the contrastive task, the predictive task and the task of distilling soft targets are respectively detached. We report their performances on the two datasets in Figure 3. Here we adopt the settings of $\textbf{Stu-1}$ in Table 4 for its best performance on the two datasets. 
\par

By associating Table 3 and and Fig. 3, we can observe that, the student model almost outperforms teacher model on the two datasets, showing the great effectiveness of our proposed framework, which is out of our expectation. Actually, similar results are also reported in other related research \cite{wu2020lite}. It indicates there is much parameter redundancy in Transformer-based recommendation models. In addition, though with a tiny size, $\textbf{Stu-base}$ even shows better performance than STAMP and SR-GNN in most cases, implying that our model is an ideal solution for on-device recommendation. By comparing $\textbf{Stu-base}$ with $\textbf{Stu-TTD}$, we can draw a conclusion that the semi-tensor product can further compress the model without any accuracy sacrifice. Comparing $\textbf{Stu-base}$ with the full version of the student, KD improves the model by 16.86\% on P@5, 12.31\% on M@5, 16.35\% on P@10, and 12.34\%
on M@10 on Tmall and improves the model by 9.53\% on P@5, 3.21\% on M@5, 11.24\% on P@10 and 3.58\% on M@10 on RetailRocket. It is concluded that the proposed self-supervised KD framework creates effective self-supervised signals and successfully transfers rich knowledge from the teacher to the student. As for the specific distillation tasks, we can see that all the three tasks are helpful. The task of distilling soft targets contributes the most among the three, while the contributions of the contrastive and predictive tasks are not negligible. \par

\begin{table}[h]
	\small
	\renewcommand\arraystretch{1.0}
	\label{Table:7}
	\begin{center}
		\begin{tabular}{cccccc}
			\hline
			time &  GRU4Rec & NARM&STAMP&SR-GNN &Student   \\ \hline
			Tmall& 0.417& 0.264& 0.175&0.310 & 0.169 \\
			RetailRocket&0.093 & 0.096&0.094 &0.192 &0.081 \\
			\hline
		\end{tabular}
	\end{center}
	\caption{Prediction time (s) on device.}
	\vspace{-10pt}
\end{table}

\subsection{Efficiency Comparison}
The low latency is a prominent advantage of on-device models. We also study the efficiency of the student model on resource-constrained devices. We first train the student model on GPU with the help of the teacher, and then the well-trained student is encapsulated by PyTorch Mobile. We use Android Studio 11.0.11 to simulate virtual device environment and deploy the student on this virtual environment to do the inference. The selected device system is Google Pixel 2 API 29. 
We compare the prediction time for every 100 sessions of each model on the two datasets. For a fair comparison, these baselines are also deployed on the same virtual environment. The results are reported in Table 7. Obviously, our student model is faster than all the baselines with a much tinier model size on the two datasets. 

\subsection{Performances on Long-Tail Items}
To relieve the data sparsity problem, we propose to categorize the learned representations into the hot and cold and then swap and recombine their session representations. To verify whether the proposed KD can address this issue, we split test set into two subsets based on the popularity of the ground-truth items. 95\% items with the fewest clicks are labeled `long-tail', and the left 5\% items with the most clicks are labeled as `Popular'. Then we compare the proportion that each category of items contributes to Pre@5 of the student model with or without self-supervised knowledge distillation framework on the two datasets. The results are presented in Table 8. On both datasets, we observe that after applying our proposed self-supervised knowledge distillation, there are growths on Pre@5 and the proportion contributed by the long-tail items significantly increases. It implies that the proposed KD framework can effectively relieve the data sparsity problem and the embedding recombination strategy makes student learn more expressive representations under the direct help of the teacher.
\begin{table}[h]
	\small
		\label{Table:7}
		\begin{center}
		\setlength{\tabcolsep}{1mm}{{{
			\begin{tabular}{*{5}{c}}
				\toprule
				\multirow{2}{*}{Model} &
				\multicolumn{2}{c}{Tmall} & \multicolumn{2}{c}{RetailRocket} \cr
				\cmidrule(lr){2-3}\cmidrule(lr){4-5} & Popular & Long-tail & Popular & Long-tail\\ \hline
				Stu-base & 5.30 & 14.93 &7.50 & 24.61 \\
				Student& 6.03 & 17.43 &7.72 & 27.45 \\
				\hline
			\end{tabular}}}}
		\end{center}
		\caption{Performance on long-tail items.}
		\vspace{-15pt}
	\end{table}

\subsection{Hyperparameter Analysis}
We explore the sensitiveness of the three coefficients used in Eq. (21) on the two datasets. A large number of values are searched and we choose to report some representative values of the three parameters; they are \{0.005, 0.01, 0.02, 0.05, 0.1, 0.2, 0.5\} for $\beta_1$, \{0.001, 0.005, 0.01, 0.02, 0.05, 0.1, 0.2\} for $\beta_2$ and \{0.2, 0.4, 0.5, 0.6, 0.7, 0.8, 0.9\} for $\beta_3$. When investigating one hyperparameter of them, we fix the rest at the best values of them. For Tmall, the best group is (0.1, 0.001, 0.8). For RetailRocket, the best combo is (0.005, 0.1, 0.8). The Pre@5 and Pre@10 results are presented in Figure 4. Obviously, for most cases on both datasets, with the increase of each parameter, the performance firstly drops, then starts to increase until reaching the peak, and then gradually decreases. Tmall has a little different curves towards $\beta_2$, where the best performance is achieved at the starting point. The curves of $\beta_3$ on RetailRocket also show a different pattern, where with the increase of $\beta_3$, the precision values start to grow steadily until reaching the peak and then decline. Besides, for both datasets, it should be noted that the best value of $\beta_3$ is larger than $\beta_1$ and $\beta_2$, indicating that the soft targets distillation plays a more important role in improving the student model, which is in line with the results in Figure 3.


\begin{figure}[t]  
    \centering
	\vspace{-10pt}
    \subfloat[Tmall]{%
      \includegraphics[width=0.5\textwidth]{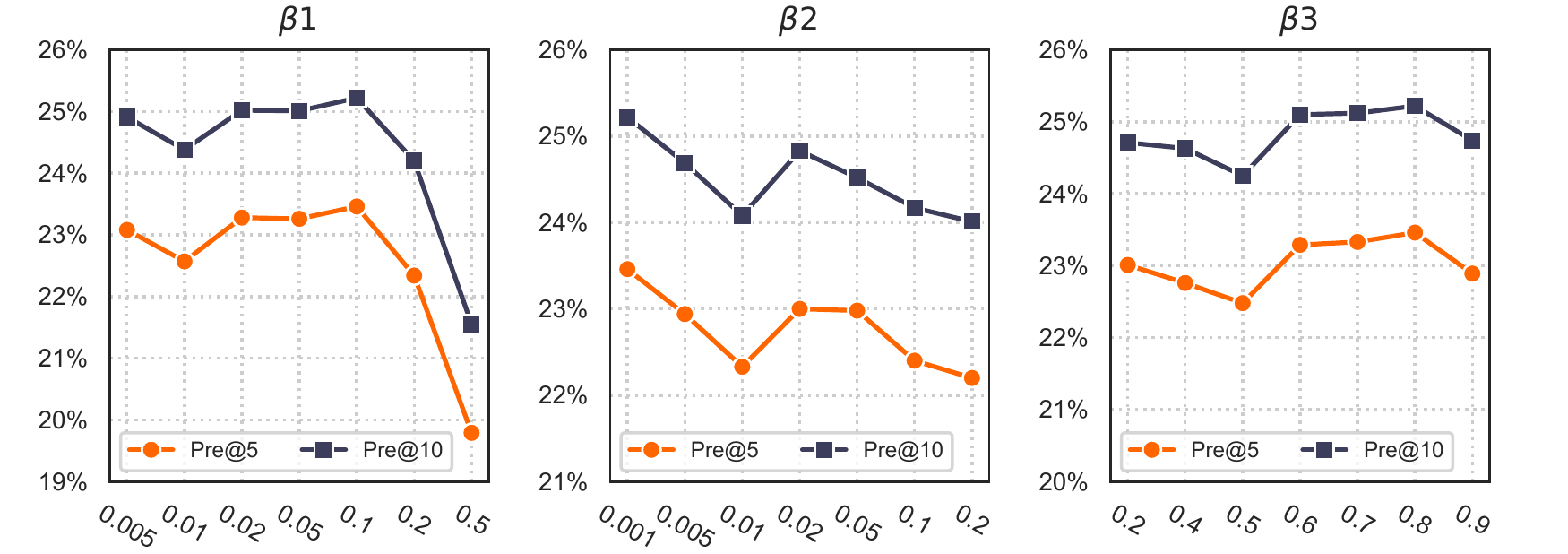}%
    }   
    \\
    \subfloat[RetailRocket]{%
      \includegraphics[width=0.5\textwidth]{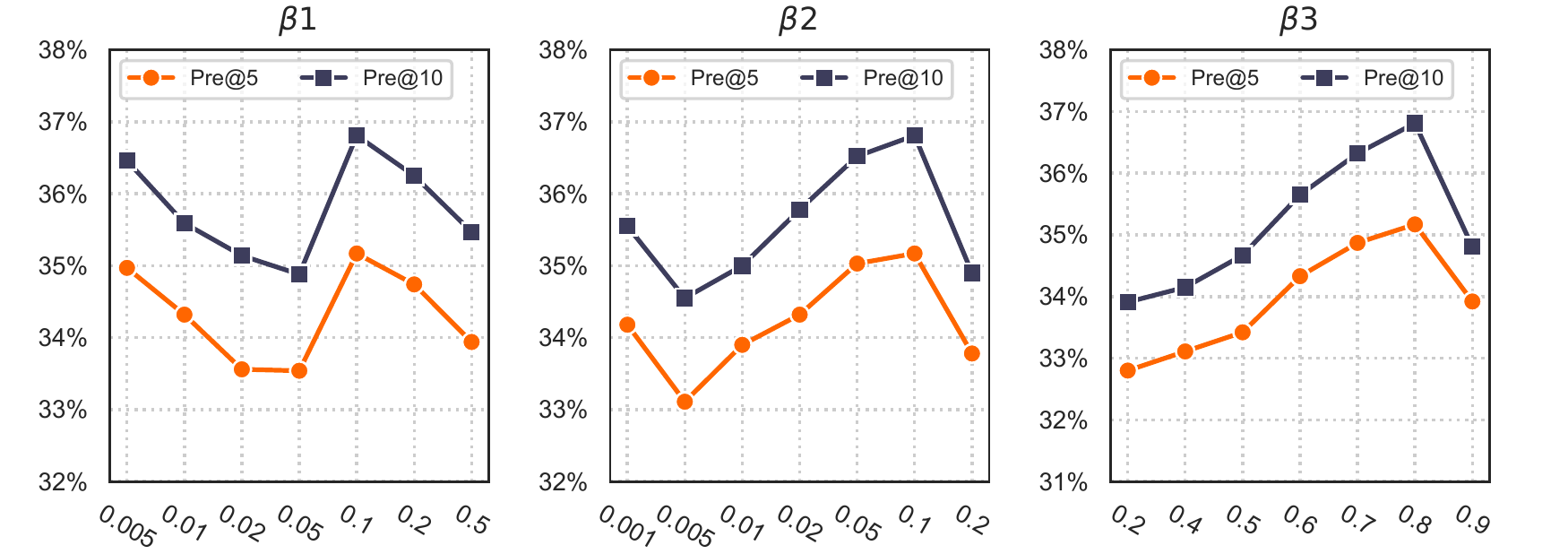}%
    }   
    \caption{Hyperparameter Analysis.}
    \label{figure:4} 
	\vspace{-10pt}
\end{figure}


\section{Conclusion}
Current recommender systems work in a cloud-based paradigm where models are trained and deployed in the server and use abundant memory storage and computing resources to generate recommendation. However, this scheme raises the privacy concern among users and it is not energy-saving. On-device recommendation addresses these issues by deploying lightweight models on edge devices and perform local inference. In this paper, we propose a novel on-device recommendation model for next-item recommendation in the session-based scenario. We loose the dimension consistency constraint in tensor decomposition for an ultra-compact model and propose a self-supervised knowledge distillation framework to compensate for the capacity loss caused by the compression. Extensive experimental results demonstrate that, with a 30x size reduction, our proposed model can achieve comparable and even higher performances compared with its regular counterpart.

\section{Acknowledgement}
This work is supported by Australian Research Council Future
Fellowship (Grant No. FT210100624), Discovery Project (Grant No.
DP190101985) and Discovery Early Career Research Award (Grant
No. DE200101465).
\bibliographystyle{ACM-Reference-Format}
\bibliography{ref}
\end{document}